\documentclass[preprint, showpacs, aps, draft]{revtex4}
\usepackage{bm}

\begin{document}
\title{Interacting classical and quantum ensembles}
\author{Michael J. W. Hall}
\affiliation{Theoretical Physics, IAS, \\ Australian National
University,\\
Canberra ACT 0200, Australia}

\author{Marcel Reginatto}
\affiliation{Physikalisch-Technische Bundesanstalt\\ Bundesallee 100
\\38116 Braunschweig, Germany}

\begin{abstract}
A consistent description of interactions between classical and quantum systems is relevant to quantum measurement theory, and to calculations in quantum chemistry and quantum gravity.  A solution is offered here to this longstanding problem, based on a universally-applicable formalism for ensembles on configuration space. This approach overcomes difficulties arising in previous attempts, and in particular allows for backreaction on the classical ensemble, conservation of probability and energy, and the correct classical equations of motion in the limit of no interaction.  Applications include automatic decoherence for quantum ensembles interacting with classical measurement apparatuses; a generalisation of coherent states to hybrid harmonic oscillators; and an equation for describing the interaction of quantum matter fields with classical gravity, that implies the radius of a Robertson-Walker 
universe with a quantum massive scalar field can be sharply defined only for particular `quantized' values.  
\end{abstract} 

\pacs{03.65.Ta, 04.62.+v}
\maketitle

\section{Introduction}

A fully consistent and general approach to describing interactions between classical and quantum systems is of great interest in quantum measurement theory, and is also highly relevant to calculations in quantum chemistry and quantum gravity. In particular, in the standard Copenhagen interpretation of quantum mechanics, it has been emphasised that the measuring apparatus must be described in classical terms 
\cite{bohr, heis}.  Moreover, it is often both physically appropriate and computationally convenient to assume that some components of physical systems - such as atomic nuclei, the electromagnetic field or the spacetime metric - can be modelled classically \cite{boucher, meanfield, hawking, kiefer}.  

However, despite many attempts, fundamental difficulties have always arisen in trying to implement such a `mixed' description in a consistent manner.  These difficulties include the nonconservation of energy; absence of backreaction of the quantum system on the classical system; nonlocality; negative probabilities;  an inherent inability to describe all interactions of interest; and incorrect equations of motion in the limit of no interaction.  Thus, it has not been clear as to whether a satisfactory description is even possible.

Previous attempts in the literature fall into three main categories, briefly summarised here.  

First, in the {\it mean-field} (or Born-Oppenheimer) approach, classical observables appear as parameters in a quantum Hamiltonian operator.  This operator directly specifies the evolution of the quantum system in the usual way, while its average over the quantum degrees of freedom specifies a classical Hamiltonian for the classical parameters \cite{meanfield, kiefer}.  However, while computationally convenient, such an approach cannot couple any statistical fluctuations into the classical observables, which are required, for example, if measurement interactions are to lead to a multiplicity of possible outcomes \cite{boucher}. Diosi and Halliwell have therefore proposed a modification in which white noise is added to the mean-field evolution of the classical system, while the quantum evolution is modified by a corresponding noise-dependent nonlinear term \cite{diosihalli}.  However, the proposed model is non-generic and does not conserve energy.  In the gravitational context, stochastic semiclassical gravity provides a formally similar modification. The semiclassical Einstein equation is replaced by an Einstein Langevin equation that models the effect of quantum fluctuations of the stress energy operator that acts as a source of the gravitational field \cite{hu}. The aim of the theory is to provide an intermediate step between semiclassical and quantum gravity, one which takes into account quantum fluctuations and metric perturbations to first order. It is therefore not intended to provide a complete description of quantum matter fields interacting with a classical gravitational field.

Second, the {\it phase space} (or algebraic) approach relies on modelling the classical system by a set of mutually commuting `phase space' observables on some Hilbert space, and allowing a unitary interaction with the quantum system.  The most sophisticated model of this type is by Sudarshan and co-workers \cite{sudpramana, sud1, sud2, sud3}, in which the interaction Hamiltonian depends on non-observable operators associated with the classical system.  However, while this model has many interesting properties, it is only self-consistent for a certain class of interactions (that ensure the classical observables remain `classical'), and cannot describe, for example, the standard Stern-Gerlach measurement interaction \cite{sud2}. Moreover, a number of counterexamples and no-go theorems show that other proposed types of interaction lead to at least one of the following problems: negative probabilities; the absence of any backreaction on the classical system from the interaction; or to the loss of the correspondence principle in the classical limit 
\cite{boucher, salcedo1, salcedo2, royal, ternoperes, sahoo}.  

Diosi et al. have proposed a variation on the phase space approach, in which the classical phase space is mapped to a family of coherent states on Hilbert space rather than to a set of orthonormal states \cite{royal}.  However, while this is successful in ensuring positive probabilities and incorporating backreaction, it yields incorrect equations of motion for the classical system in the limit of no interaction; does not permit a well-defined position or momentum to be ascribed to a classical particle at {\it any} time (not even initially); and is intrinsically ambiguous in that the family of coherent states is only specified up to an arbitrary parameter $L$ (corresponding to the choice of length $L$ making the combination $a=(x/L + iLp/\hbar)/\sqrt{2}$ dimensionless), on which the equations of motion are dependent.  An alternative variation by Dias and Prata couples the quantum system to a {\it quasi}classical system 
\cite{dias}, but by construction yields `quasiclassical' equations of motion rather than the usual classical equations of motion in the limit of no interaction.

Third and finally, the {\it trajectory} approach relies on modifying the deBroglie-Bohm formulation of quantum mechanics, wherein quantum states correspond to an ensemble of  trajectories acted on by a `quantum potential' \cite{bohm}.  In particular, the equations of motion for these trajectories can be modified in various ad hoc ways to incorporate interaction with a classical particle \cite{traj}.  Such approaches incorporate backreaction on the classical particle and have been found computationally useful for making semiclassical calculations in quantum chemistry.  However, they are generally unsuitable (and are not intended) for providing a fundamental description of classical-quantum systems; automatically inherit the various difficulties of the deBroglie-Bohm formulation \cite{dbbprobs}; and do not conserve energy 
\cite{salcedocom,prezreply}.

A new approach is presented in this paper, based on a canonical formalism for describing statistical ensembles on configuration space \cite{super}.  This formalism is applicable to both classical and quantum ensembles, and is able to fully and consistently describe interactions between such ensembles. The resulting `configuration ensemble' approach is distinguished from the above mean-field and phase space approaches by the use of configuration space rather than phase space, and from the mean-field and trajectory approaches by the treatment of the classical and quantum components on an equal statistical footing.  These differences allow all of the above-identified problems associated with previous approaches to be avoided. 

Note that an apparent `incompleteness' in describing physical systems by ensembles on configuration space is that information relating to individual trajectories is simply not available in general, which is not the usual picture for classical particles in particular.  However, it will be shown that there is in fact no need to supplement the ensemble description (although this remains an option): a trajectory picture can be recovered for ensembles of classical particles precisely in those cases that trajectories are operationally defined (i.e., via the results of consecutive position measurements separated by small time intervals).  Hence the description is physically complete.  

The description of ensembles on configuration space is reviewed in Sec.~II.A, and is shown to incorporate the case of interacting quantum and classical ensembles in Sec.~II.B.  Some general features of these quantum-classical ensembles (or hybrid systems) are briefly noted in Sec.~II.C, including conditional wavefunctions, Galilean invariance, stationary states, and the nonseparability of centre-of-mass and relative motions for central forces. 

The remainder of the paper is directed to several quite diverse applications of interest.  Thus, it is shown in Sec.~III that the conditional wavefunction for the quantum component of a composite ensemble automatically decoheres under measurement-type interactions with the classical component. This yields a possible solution to the quantum measurement problem, via the assumptions that (i) measuring devices must be described classically, and (ii) the configuration of the measuring device is an operationally well-defined quantity that provides the only information available to us regarding the configuration of the quantum system.

Classical and quantum harmonic oscillators have long been of interest as physical models, and hence some aspects of the hybrid harmonic oscillator, corresponding to joining a classical particle to a quantum particle via a spring, are investigated in Sec.~IV.  In particular, it is shown that `coherent states' may be defined for such an oscillator, with properties intermediate to the fully classical and the fully quantum cases.

In Sec.~V a consistent equation is written down for describing the interaction of quantum matter fields with classical spacetime.  Assuming the configuration of the gravitational metric to be operationally well-defined leads to gravitationally-induced decoherence of the quantum state of matter.  Solutions are investigated for the special case of a Robertson-Walker spacetime interacting with a scalar field on minisuperspace, for which it is shown that solutions having a well-defined radius $a$ of the universe are only permitted for certain `quantized' 
values of $a$.  Further, these solutions 
may throw some light on the well known `arrow of time' problem \cite{kiefer}.

Conclusions are given in section VI.

\section{Configuration-space ensembles}

\subsection{Formalism}

The description of physical systems by ensembles on configuration space may be introduced at quite a fundamental and generic level, requiring little more than the notions of probability and an action principle \cite{super}.  

Suppose first, as seems to be implied by quantum mechanics, that the configuration of a physical system is an inherently statistical concept.  The system must therefore be described by an ensemble of configurations, corresponding to some probability density $P$ on the configuration space.  

For example, for a continuous configuration space corresponding to the possible positions of a point particle, the ensemble at time $t$ must be described by a continous probability density $P(x,t)$, with
\[ \int dx \,P(x,t) = 1 .\]
Similarly, for a discrete configuration space corresponding to the results of a coin toss or to the energy levels of a bound electron, the ensemble must be described by a discrete probability distribution $P_j(t)$, with
\[ \sum_j \,P_j(t) =1 . \]
Likewise, for a configuration space comprising a set of functions, such as the configurations of a scalar field, the ensemble is described by a probablility density functional $P[f,t]$, with
\[ \int Df \, P[f,t] =1 , \]
where $Df$ denotes the functional integration measure.

Second, suppose that the dynamics of the ensemble is specified by an action principle.  In the Hamiltonian formalism, this implies the existence of a function $S$ on configuration space that is canonically conjugate to $P$, and an {\it ensemble} Hamiltonian $\tilde{H}[P,S]$ satisfying the action principle $\delta A = 0$, with \cite{goldstein}
\begin{equation} \label{action}
A = \int dt\, \left[ - \tilde{H} + \int_{c.s.} S\frac{\partial P}{\partial t} \right] \simeq - \int dt\, \left[ \tilde{H} + \int_{c.s.} P \frac{\partial S}{\partial t} \right] ,
\end{equation}
where $\simeq$ denotes equality up to a total time derivative (which does not affect the equations of motion \cite{goldstein}).  Here $\int_{c.s.}$ is used to denote integration over the configuration space, and is replaced by summation for any discrete parts of the configuration space.

Thus, for example, the equations of motion for $P$ and $S$ in the case of a continuous configuration space follow via Eq.~(\ref{action}) as \cite{goldstein}
\begin{equation} \label{motioncont}
\frac{\partial P}{\partial t} = \frac{\delta\tilde{H}}{\delta S}, ~~~~~~~ \frac{\partial S}{\partial t} = -\frac{\delta\tilde{H}}{\delta P},
\end{equation}
where $\delta/\delta f$ denotes the usual functional derivative (in particular, for $F[f] = \int dx\,R(f,\nabla f)$, one has $\delta F/\delta f = \partial R/\partial f - \nabla .(\partial R/\partial \nabla f)$).  Similarly, for the case of a discrete configuration space, the equations of motion for the ensemble follow from Eq.~(\ref{action}) as 
\begin{equation} \label{motiondisc}
\frac{\partial P_j}{\partial t} = \frac{\partial \tilde{H}}{\partial S_j},~~~~~~ \frac{\partial S_j}{\partial t} = -\frac{\partial 
\tilde{H}}{\partial P_j} .
\end{equation}

Eq.~(\ref{action}) above encapsulates the basic formal content of the configuration ensemble formalism.  Some general properties have been summarised elsewhere \cite{super}.  For example, it may be shown that conservation of probability implies the constraint that any physical ensemble Hamiltonian is invariant under the addition of an global constant to $S$, while positivity of probability at all times implies that $\partial P/\partial t$ vanishes whenever $P$ does.  Thus,
\begin{equation} \label{invar}
\tilde{H}[P,S+c] = \tilde{H}[P, S],~~~~~~~\left. \frac{\delta \tilde{H}}{\delta S}\right|_{P(x)=0} = 0,
\end{equation}
for any constant $c$, where $\delta/\delta S$ is replaced by $\partial/\partial S_j$ for discrete configuration spaces.  It follows in particular that only relative values and derivatives of $S$ have dynamical significance, i.e., $S$ is physically well-defined only up to an arbitrary additive constant.  

Two ensemble Hamiltonians of particular interest are 
\begin{eqnarray} \label{hc}
\tilde{H}_C[P,S] &=& \int dx\, P \left[ \frac{|\nabla S|^2}{2m} + V(x)\right] ,\\ \label{hq}
\tilde{H}_Q[P,S] &=& \tilde{H}_C[P,S] + \frac{\hbar^2}{4} \int dx\, P\frac{|\nabla \log P|^2}{2m} .
\end{eqnarray}
The equations of motion for $\tilde{H}_C$ follow from 
Eq.~(\ref{motioncont}) as
\begin{equation} \label{class}
\frac{\partial P}{\partial t} + \nabla .\left( P\frac{\nabla S}{m} \right) =0,~~~~~~\frac{\partial S}{\partial t} + \frac{|\nabla S|^2}{2m} + V = 0, \end{equation}
which may be recognised as the continuity equation and Hamilton-Jacobi equation for an ensemble of {\it classical} particles of mass $m$ moving in a potential $V(x)$ \cite{goldstein}.  In contrast, the equations of motion for $\tilde{H}_Q$ follow from Eq.~(\ref{motioncont}) as being equivalent to the real and imaginary parts of the Schr\"{o}dinger equation
\begin{equation} \label{se}
i\hbar \frac{\partial \psi}{\partial t} = \frac{-\hbar^2}{2m}\nabla^2\psi + V\psi ,\end{equation}
where $\psi:=P^{1/2}e^{iS/\hbar}$, and hence $\tilde{H}_Q$ describes an ensemble of {\it quantum} particles moving in a potential $V(x)$.  

It is seen that, in the configuration ensemble formalism, classical and quantum particles are treated on an equal footing, with differences being primarily due to the different forms of the respective ensemble Hamiltonians \cite{constraint}.  The generic form of the additional kinetic term in $\tilde{H}_Q$ is related to the Fisher information matrix $I[P]=\langle F[P]\rangle$ of classical statistics 
\cite{frieden, regpra}, where
\[ F[P] = (\nabla \log P)\,(\nabla \log P)^T , \]
and may be derived via an `exact uncertainty principle' for both particles and bosonic fields \cite{hrjpa,hkr}.  In particular, for any `classical' ensemble Hamiltonian of the form $\tilde{H}_C=\langle {\rm tr}[K\nabla S(\nabla S)^T] + V\rangle$, for some kinetic matrix function $K$ on the configuration space (thus $K=I/(2m)$ in Eq.~(\ref{hc}) above), one may derive the corresponding `quantum' form $\tilde{H}_Q=\tilde{H}_C+(\hbar^2/4)\langle {\rm tr}[KF]\rangle$.  Note that the positivity of $F[P]$ implies that the quantum ensemble energy $\tilde{H}_{Q}$ is never less than the corresponding classical ensemble energy $\tilde{H}_{C}$, as long as the kinetic matrix $K$ is also positive (the one notable exception is gravity, discussed in section V, where $K$ corresponds to the DeWitt supermetric).

The configuration ensemble formalism is very general, and may in fact be used to describe {\it all} quantum systems (eg, qubits, particles with spin, and  fermionic fields).  In particular, for any quantum ensemble described by some ket $|\psi\rangle$ and Hamiltonian operator $\hat{H}$, let $\{|a\rangle\}$ denote a complete set of kets, defining a `configuration' representation.  The ensemble Hamiltonian given by
\begin{equation} \label{hgen} 
\tilde{H}[P,S] = \langle\psi|\hat{H}|\psi\rangle , 
\end{equation}
with $P$ and $S$ defined via the polar decomposition $\langle a|\psi\rangle =P^{1/2}e^{iS/\hbar}$, then yields equations of motion equivalent to the Schr\"{o}dinger equation $i\hbar\partial|\psi\rangle/\partial t=\hat{H}|\psi\rangle$ in the $\{|a\rangle\}$ representation 
\cite{super} (indeed, $\psi(a)=\langle a|\psi\rangle$ and its complex conjugate $\psi^*(a)$, appearing in the standard Hamiltonian formulation of the Schr\"{o}dinger equation \cite{roman}, are related to $P$ and $S$ by a canonical transformation).  The ensemble Hamiltonian 
$\tilde{H}_Q$ in Eq.~(\ref{hq}) represents a special case of this formula. 

It is important to note that, for the configuration ensemble formalism to maintain full generality across the classical and quantum spectrum, no limiting interpretation should be assigned to $S$ (although 
it may be shown for continuous configuration spaces that $\langle \partial_\mu S\rangle$ is related to the ensemble energy-momentum in many cases of interest \cite{super}).  Thus, $S$ will be regarded here simply as the canonical conjugate of the probability density $P$, with its existence being an immediate consequence of the requirement of an action principle for $P$.  

In particular, for an ensemble of classical particles, described by Eqs.~(\ref{hc}) and (\ref{class}), it will {\it not} be assumed that the {\it velocity} of a member of the ensemble at position $x$ is a physically well-defined quantity given by $\nabla S/m$, contrary to the usual trajectory interpretation of the Hamilton-Jacobi equation 
\cite{goldstein}.  This avoids forcing a similar trajectory interpretation in the quantum and quantum-classical cases (although this remains as an option if desired, albeit attended by the types of difficulties associated with the deBroglie-Bohm approach 
\cite{dbbprobs}).  Such an assumption is in fact {\it unnecessary} for classical ensembles:  if an ensemble is well-localised about some position $x_0$ at time $t=0$ (eg, via a position measurement at time $t=0$), such that $\nabla S\approx mv$ over the support of $P$ for some constant vector $v$, then the ensemble will be well-localised about $x_0+vt$ a sufficiently short time $t$ later as a direct result of Eqs.~(\ref{class}).  Hence, dividing the difference of two successive position measurements by $t$ will give a result close to $v$, {\it independently of any assumptions about underlying trajectories}.  

More precisely, if the effective width of $P(x,0)$ following a first position measurement at time $t=0$ is sufficiently small, then $\nabla S$ can be taken to be approximately constant over the support of the ensemble, so that $\nabla S(x,0)\approx mv$ with $\nabla .v=0$. The continuity equation and the gradient of the Hamilton-Jacobi equation in Eq.~(\ref{class}) are then approximated to the same level of accuracy by
\[ \frac{\partial P}{\partial t} + v .\nabla P \approx 0,~~~~~~m\frac{dv}{d t} + \nabla V \approx 0 , \]
and the probability density at time $t$ therefore follows as
\[ P(x,t) \approx P(x-x_t,0) , \]
where $x_t:=\int_0^t ds\, v(s)$ is the solution of Newton's equation $md^2x_t/dt^2=mdv/dt=-\nabla V$.  Thus, if $P(x,0)$ is well-localised about $x_0$, a second measurement of position at time $t$ will be well-localised about $x_0+x_t$ with high probability (up to times $t$ for which the approximation that $\nabla S$ is spatially uniform over the width of the ensemble remains valid).  This point, that {\it no underlying trajectory interpretation for classical ensembles is actually required for describing the trajectory-like behaviour of successive position measurements}, has also been emphasised recently by Nikolic \cite{nik}.  It is also reminiscent of the case in hydrodynamics, where the flowlines do not represent the trajectories of the molecules comprising the fluid being described.

Finally, it is worth noting that there is a corresponding Lagrangian formulation of the configuration ensemble formalism, based on the Lagrangian
\[ \tilde{L}[P,\partial P/\partial t,S,\partial S/\partial t] := -\tilde{H}[P,S] + \int_{c.s.} S\frac{\partial P}{\partial t} \simeq -\tilde{H}[P,S] - \int_{c.s.} P\frac{\partial S}{\partial t}   \]
(where $\simeq$ denotes equality up to a total time derivative).  Clearly the associated action $\int dt\,\tilde{L}$ is equal to $A$ in Eq.~(\ref{action}), and hence yields the same equations of motion.
However, note that (essentially as a consequence of the action being at most linear in the time derivatives of $P$ and $S$), the Lagrangian formulation requires the {\it ab initio} introduction of two independent functions $P$ and $S$ on configuration space, leading further to two conjugate quantities $\pi_P=\delta\tilde{L}/\delta (\partial P/\partial t)$, $\pi_S=\delta\tilde{L}/\delta (\partial S/\partial t)$.  In constrast, the direct Hamiltonian formulation only requires the introduction of a single function $P$ on configuration space, with $S$ then being {\it defined} as the single canonically conjugate quantity.  Hence the Hamiltonian formulation is simpler, and so may be considered more fundamental.  

\subsection{Interacting ensembles}

It has been shown that the configuration ensemble formalism encompasses the descriptions of both standard quantum and classical ensembles, making it a suitable starting point for considering interactions between such ensembles.  

Now, if $A$ and $B$ denote two general configuration spaces, the joint configuration space is just the set product $A\times B$.  A composite ensemble is therefore described by a joint probability density $P(a,b)$ on $A\times B$, a conjugate quantity $S(a,b)$, and a {\it composite} ensemble Hamiltonian $\tilde{H}_{AB}[P,S]$. It is actually quite straightforward to write down composite ensemble Hamiltonians $\tilde{H}_{AB}$ that correspond to interacting quantum and classical ensembles (see, eg, Eqs.~(\ref{vqx}) and (\ref{hspin}) below).  
However, it is of interest to first briefly discuss just what is meant by `interacting' and `non-interacting' ensembles in general.

In particular, consider two ensembles defined on configuration spaces $A$ and $B$ respectively, that are {\it independent} at some given time, i.e., the composite ensemble is fully described by the conjugate pairs $P_A(a)$, $S_A(a)$ and $P_B(b)$, $S_B(b)$ on $A$ and $B$ respectively.  Hence no physical distinction is possible between the joint probability density $P(a,b)$ assigned to the composite ensemble, and the pair of individual densities $P_A(a)$ and $P_B(b)$, implying from basic probability theory that
\begin{equation} \label{pind} P(a,b) = P_A(a)\,P_B(b) . \end{equation}
Similarly, there can be no physical distinction between the joint function $S(a,b)$ and the pair of individual functions $S_A(a)$ and $S_B(b)$.  Recalling that conservation of probability requires that such functions are only physically well-defined modulo an additive constant on their respective configuration spaces, as per Eq.~(\ref{invar}), it follows that $S(a,b)$ must be equivalent to $S_A(a)$ up to some additive function of $b$, and be equivalent to $S_B(b)$ up to some additive function of $a$, and hence that
\begin{equation}\label{sind} S(a,b) = S_A(a) + S_B(b)   \end{equation}
(up to an arbitary additive constant of no physical significance).  Note for quantum ensembles that independence thus corresponds to factorisability of the joint wavefunction $\psi(a,b)$.

A composite ensemble Hamiltonian $\tilde{H}_{AB}$ may now be defined as describing {\it non-interacting} ensembles if and only if {\it all initially independent ensembles remain independent under evolution}.  It follows from the form of the ensemble action in Eq.~(\ref{action}) that this is equivalent to the condition 
\begin{equation} \label{nonint} 
\tilde{H}_{AB}[P_AP_B, S_A+S_B] \simeq \tilde{H}_A[P_A,S_A] + \tilde{H}_B [P_B,S_B] 
\end{equation}
for all $P_A(a)$, $P_B(b)$, $S_A(a)$ and $S_B(b)$, where $\tilde{H}_A$ and $\tilde{H}_B$ are ensemble Hamiltonians on $A$ and $B$ respectively (and where $\simeq$ denotes equality up to a total time derivative).  If Eq.~(\ref{nonint}) is {\it not} satisfied, $\tilde{H}_{AB}$ will be said to describe {\it interacting} ensembles \cite{interact}.  

It follows, for example, that any pair of ensemble Hamiltonians of the forms
\begin{eqnarray} \label{forma} \tilde{H}_A(P_A,S_A) &=& \int da \,P_A\, F(\nabla_a \log P_A, \nabla_a S_A,\dots),\\ \label{formb}
\tilde{H}_B(P_B,S_B) &=& \int db \,P_B\, G(\nabla_b \log P_B, \nabla_b S_B,\dots) ,
\end{eqnarray}
for two continuous configuration spaces $A$ and $B$ and arbitary functions $F$ and $G$ (where `$\dots$' denotes possible higher derivatives of $\log P$ and $S$), can be extended to the non-interacting composite ensemble Hamiltonian
\[ \tilde{H}_{AB}[P,S] := \int da\,db\, P\, \left[  F(\nabla_a \log P,\nabla_a S,\dots) + G(\nabla_b \log P,\nabla_b S,\dots) \right]  \]
satisfying Eq.~(\ref{nonint}).  In particular, since the ensemble Hamiltonians $\tilde{H}_C$ and $\tilde{H}_Q$ in Eqs.~(\ref{hc}) and (\ref{hq}) are of the forms of Eqs.~(\ref{forma}) and (\ref{formb}), they can be extended to ensemble Hamiltonians corresponding to any one of non-interacting classical-classical, quantum-quantum, and quantum-classical ensembles of particles.

As an explicit example of {\it interacting} quantum and classical ensembles, let $q$ denote the configuration space coordinate of a quantum particle of mass $m$, and $x$ denote the configuration coordinate of a classical particle of mass $M$.  A composite ensemble Hamiltonian, corresponding to some `interaction potential' $V(q,x,t)$, is then given by
\begin{equation} \label{vqx}
\tilde{H}_{QC}[P,S] = \int dq\,dx\, P\,\left[ \frac{|\nabla_q S|^2}{2m}
+ \frac{|\nabla_x S|^2}{2M} + V(q,x,t) + \frac{\hbar^2}{4} \frac{|\nabla_q \log P|^2}{2m} \right] .
\end{equation}
For $V(q,x)\equiv 0$ there is no interaction between the quantum and classical parts of the composite ensemble, and Eq.~(\ref{nonint}) is satisfied, with $\tilde{H}_{QC}$ corresponding to the sum of $\tilde{H}_Q$ and $\tilde{H}_C$ in Eqs.~(\ref{hc}) and (\ref{hq}).  More generally, $\tilde{H}_{QC}$ is seen to correspond to the sum of a quantum term, a classical term, and and an interaction term
\[ \tilde{H}_I[P,S] = \int dq\,dx\, P\,V(q,x,t)= \langle V\rangle  .\]

The equations of motion corresponding to Eq.~(\ref{vqx}) follow via 
Eq.~(\ref{motioncont}) as 
\begin{equation} \label{cont}
\frac{\partial P}{\partial t} =  -\nabla_q .\left( P 
\frac{\nabla_qS}{m} \right) - \nabla_x.\left(P\frac{\nabla_xS}{M}\right) ,
\end{equation} 
\begin{equation} \label{hj}
\frac{\partial S}{\partial t} = - 
\frac{|\nabla_qS|^2}{2m} - \frac{|\nabla_xS|^2}{2M} - V +
\frac{\hbar^2}{2m}\frac{\nabla_q^2 P^{1/2}}{P^{1/2}} .
\end{equation}
These are coupled partial differential equations of first-order in time (of a type commonly encountered in hydrodynamics), and may be numerically integrated to solve for $P(q,x,t)$ and $S(q,x,t)$ providing that $P$ and $S$ are specified at some initial time $t_0$.  In many cases of interest the classical and quantum ensembles will be initially independent, implying the initial forms 
\[ P(q,x,t_0) = P_Q(q)\,P_X(x),~~~~~~~S(q,x,t_0) = S_Q(q) + S_X(x) \]
for $P$ and $S$ as per Eqs.~(\ref{pind}) and (\ref{sind}).

As an example of a {\it discrete} ensemble of quantum systems interacting with an ensemble of classical particles, consider the case where the $z$-component $\hat{\sigma}_z$ of an ensemble of spin-1/2 particles is linearly coupled to the momentum of an ensemble of one-dimensional classical particles.  Modelling such an interaction has been attempted previously in the context of the phase space approach \cite{royal}, where, however, a number of fundamental difficulties arise (see Sec.~I above).  The simplest way to proceed in the configuration ensemble formalism is to (i) temporarily `promote' the classical particles to quantum particles; (ii) write down the corresponding quantum-quantum composite ensemble Hamiltonian using the prescription in Eq.~(\ref{hgen}); and then (iii) take the classical limit for the promoted quantum component to obtain the desired hybrid ensemble Hamiltonian.

In particular, the Hamiltonian operator describing a linear coupling between the momentum $\hat{p}$ of a {\it quantum} one-dimensional particle of mass $M$ and the component $\hat{\sigma}_z$ of a quantum spin-1/2 system has the form
\begin{equation} \label{hamop}  
\hat{H} = \frac{\hat{p}^2}{2M} + V(\hat{x}) + \kappa(t)\hat{p}
\, \hat{\sigma}_z  .
\end{equation}
Choosing a basis $\{ |x,\pm\rangle\}$, diagonal with respect to the position operator $\hat{x}$ and the spin component $\hat{\sigma}_z$, i.e., 
\[  \hat{x}\,\hat{\sigma}_z |x,\pm\rangle = \pm x\, |x,\pm\rangle ,\]
and writing $\psi(x,\pm)=\langle x,\pm|\psi\rangle = P^{1/2}e^{iS/\hbar}$, the corresponding quantum-quantum ensemble Hamiltonian can be calculated from Eqs.~(\ref{hgen}) and (\ref{hamop}) as
\begin{eqnarray*}
\tilde{H}_{QQ}[P,S] &=& \sum_{\alpha=\pm}\, \int dx \, P(x,\alpha) \left[ \frac{1}{2M}\left( 
\frac{dS(x,\alpha)}{dx}\right)^2 +V(x)\right] \\ 
~~&~&+~ \frac{\hbar^2}{4}\sum_{\alpha=\pm} \, \int dx\, P(x,\alpha) \frac{1}{2M}\left( \frac{d\log P(x,\alpha)}{dx}\right)^2 \\
~~&~& +~\kappa(t) \sum_{\alpha=\pm}\, \int dx\, \alpha P(x,\alpha)\frac{dS(x,\alpha)}{dx} .
\end{eqnarray*}
Note that $P(x,\pm)$ is the probability density for the position being $x$, and the spin being `up' or `down' in the $z$-direction.  Finally,
taking the limit $\hbar\rightarrow 0$ for the one-dimensional particle component removes the purely quantum kinetic
term on the second line (cf. Eqs.~(\ref{hc}) 
and (\ref{hq})), yielding the corresponding ensemble Hamiltonian
\begin{eqnarray} 
\tilde{H}_{\rm spin}[P,S] &=& \sum_{\alpha=\pm} \,\int dx \, P(x,\alpha) \left[ \frac{1}{2M}\left( 
\frac{dS(x,\alpha)}{dx}\right)^2 +V(x)\right] \nonumber\\ 
\label{hspin}
~~&~& +~\kappa(t)\sum_{\alpha=\pm} \,\int dx\, \alpha P(x,\alpha)\frac{dS(x,\alpha)}{dx}  
\end{eqnarray}
for a {\it classical} ensemble of particles interacting with a quantum ensemble of spin-1/2 systems.
This ensemble Hamiltonian will be further studied in Sec.~III in the context of measurement interactions.

The method used above, to obtain a hybrid ensemble Hamiltonian from an associated fully quantum ensemble Hamiltonian, is straightforward to extend to the (rather general) case where the associated Hamiltonian operator can be represented as a sum, $\hat{H}=\sum_n \hat{A}_n\otimes \hat{B}_n$, of products of operators on the respective Hilbert spaces \cite{con}.  Note that since any hybrid ensemble Hamiltonian obtained by this method is merely a particular limit of a fully quantum ensemble Hamiltonian, the hybrid evolution will preserve the positivity of $P$ as a direct consequence of it being preserved under the fully quantum evolution.

\subsection{Some general properties}

It has been shown that interacting classical and quantum ensembles are quite straightforward to discuss in the language of configuration space ensembles, requiring only the choice of a suitable composite ensemble Hamiltonian to be made.  Here some general properties of interest are briefly noted.

1. {\it Conditional wavefunctions and density operators}:  
Denoting the classical and quantum configuration variables by $x$ and $q$ respectively, the conditional probability of quantum configuration $q$, relative to a given classical configuration $x$, follows from standard probability theory as
\[  P(q|x) = P(q,x)/P(x) . \]
Here $P(x)$ denotes the marginal probablity density $\int dq\,P(q,x)$ (with integration replaced by summation over any discrete parts of the quantum configuration space).  A {\it conditional wavefunction} for the quantum component is then naturally defined by
\begin{equation} \label{cond}
\psi(q|x) := [P(q|x)]^{1/2}\, e^{i S(q,x)/\hbar} ,
\end{equation}
with corresponding ket 
$|\psi_x\rangle := \int dq\,\psi(q|x)\,|q\rangle$.
One can also define a {\it conditional density operator}
\begin{equation} \label{rho}
\rho_{Q|C} := \int dx\, P(x)\,|\psi_x\rangle\langle \psi_x| 
\end{equation}
for the quantum component (with integration again replaced by summation over any discrete parts of configuration space).  
Note, however, that $\psi(q|x)$ and $\rho_{Q|C}$ do not satisfy linear Schr\"{o}dinger and Liouville equations, nor unitary invariance properties, when there is nontrivial interaction/correlation with the classical component of the composite ensemble.  Moreover, these quantities only contain partial information about the composite ensemble: {\it a hybrid classical-quantum ensemble requires both $P(q,x)$ and $S(q,x)$ for its full description, and is not merely equivalent to some `classical' mixture of `quantum' states}. However, both $|\psi_x\rangle$ and $\rho_{Q|C}$ are useful as derived concepts, for example, in the discussion of measurement and decoherence in Sec.~III.  

2. {\it Galilean-invariant interactions}:
If $V$ in Eq.~(\ref{vqx}) transforms as a scalar under time and space translations and under rotations, i.e., 
\[ V(q,x,t)\equiv V(|q-x|) ,\] 
then it is straightforward to show that the equations of motion in 
Eqs.~(\ref{cont}) and (\ref{hj}) are invariant under the general Galilean transformation
\[ q\rightarrow q'=Rq-ut+a,~~x\rightarrow x'=Rx-ut+a,~~t\rightarrow t'=t+\tau, \]
for rotation matrix $R$, constant vectors $u$ and $a$, and constant $\tau$, provided that $P$ and $S$ transform as $P'(q',x',t') = P(q,x,t)$ and 
\[ S'(q',x',t') = S(q,x,t) + \frac{1}{2}(m+M)|u|^2t - u.R(mq+Mx)+c 
\]
for some constant $c$.  Hence, {\it Galilean invariance is satisfied for hybrid ensembles of particles}, whenever the interaction potential $V(q,x,t)$ itself is Galilean invariant.

3.  {\it Stationary states}:
For ensemble Hamiltonians with no explicit time dependence, `stationary states' may be defined as those ensembles for which the dynamical properties of the ensemble are also time-independent.  Noting that only relative values of $S$ are physically significant (see Sec.~II.A), such ensembles must satisfy the conditions $P(x,t)=P(x,t')$ and $S(x,t)-S(x',t)=S(x,t')-S(x',t')$ for all configurations $x, x'$ and times $t, t'$, which are equivalent to the conditions $\partial P/\partial t=0$ and  $\partial S/\partial t= f(t)$ for some function $f$.  Noting that 
$\dot{f}= \partial^2S/\partial t^2 = - (\partial/\partial t) (\delta \tilde{H}/\delta P)$
(where $\delta \tilde{H}/\delta P$ is replaced by $\partial \tilde{H}/\partial P_j$ for discrete configuration spaces), and that the last term must vanish if the ensemble is time-independent, it follows that {\it stationary ensembles are characterised by the conditions}
\begin{equation} \label{ss}
\frac{\partial P}{\partial t} = 0,~~~~~~~~ \frac{\partial S}{\partial t} = -E ,
\end{equation}
for some constant $E$. 

As an example, consider the case of interacting classical and quantum ensembles of particles described by the composite ensemble Hamiltonian $\tilde{H}_{QC}$ in Eq.~(\ref{vqx}), with a {\it translation-invariant} potential $V(q,x,t)\equiv V(q-x)$. Under the ansatz $P\nabla_xS=0$, it follows via Eqs.~(\ref{cont}), (\ref{hj}) and 
(\ref{ss}) that the corresponding stationary states satisfy
\[  \nabla_r.\left(P\frac{\nabla_rS}{m}\right)=0,~~~
\frac{|\nabla_rS|^2}{2m}+V(r)-\frac{\hbar^2}{2m}\frac{\nabla_r^2
\sqrt{P}}{\sqrt{P}} = E \]
(where a change of variables from $q$ to $r:=q-x$ has been made so that $\nabla_q\rightarrow \nabla_r$), which is equivalent to the 
time-independent Schr\"{o}dinger equation 
\[ \left[ \frac{-\hbar^2}{2m}\nabla_r^2 + V(r)\right]\psi = E\psi , \]
with $\psi:=P^{1/2}e^{iS/\hbar}$. The ansatz requires that $S$ does not depend on $x$, and hence (for $V\neq 0$) it must be independent of $r$. It follows that if $\phi_n(r)$ denotes any {\it real-valued} energy eigenfunction corresponding to eigenvalue $E_n$ (there is always a complete set of such eigenfunctions), there is a corresponding stationary state described by
\begin{equation} \label{stat}
P(q,x,t) = P_0(x)\, [\phi_n(q-x)]^2,~~~~~~S(q,x,t) = S_0 - E_n t ,
\end{equation}
where $P_0(x)$ is an arbitrary probability density on the classical configuration space and $S_0$ is an arbitary constant.  The corresponding numerical value of the ensemble Hamiltonian may be calculated as $\tilde{H}_{QC}=E_n$.   Further, the conditional wavefunction follows from Eqs.~(\ref{cond}) and (\ref{stat}) as the displaced energy eigenstate
\[ \psi(q|x) = \phi_n(q-x)e^{S_0-iE_nt/\hbar} . \]
Hence, {\it the stationary states have quantized energies, and correspond to eigenstates of quantum particles that are subject to the potential $V(q-x)$ with probability $P_0(x)$}.

4.  {\it Centre-of-mass and relative motion}:
While stationary states for hybrid ensembles are seen to have simple relationships to purely quantum ensembles (as do the `coherent' states studied in Sec.~IV), strikingly different relationships can arise more generally.  For example, for interacting classical and quantum ensembles of particles described by the ensemble Hamiltonian 
$\tilde{H}_{QC}$ of Eq.~(\ref{vqx}), with translationally-invariant interaction potential $V(q,x,t)\equiv V(q-x)$, consider the relabelling of the joint configuration space by the centre-of-mass and relative coordinates 
\[  \overline{x}:=\frac{mq+Mx}{m+M},~~~~~r:= q-x .\]
Rewriting Eq.~({\ref{vqx}) with respect to these coordinates, and defining the total mass $M_T$ and relative mass $\mu$ by
\begin{equation} \label{mass} 
M_T := m+M,~~~~~~\mu:= \frac{mM}{m+M},  
\end{equation}
respectively, yields
\begin{eqnarray} 
\tilde{H}_{QC} &=& \int d\overline{x}dr \, P \left[ \frac{ 
|\nabla_{\overline{x}} S|^2}{2M_T} + \frac{\hbar^2m}{4(m+M)} \frac{|\nabla_{\overline{x}} \log P|^2}{2M_T} \right]\nonumber\\
~~&~&~~ +\int d\overline{x}dr \, P \left[ \frac{
|\nabla_r S|^2}{2\mu} + \frac{\hbar^2M}{4(m+M)} \frac{|\nabla_r \log P|^2}{2\mu} + V(r) \right]\nonumber\\ \label{hrel}
~~&~&~~ - \frac{\hbar^2}{4(m+M)} \int d\overline{x}dr \, \frac{ \nabla_{\overline{x}}P.\nabla_r P}{P}  .
\end{eqnarray}
Comparing with the form of $\tilde{H}_Q$ in Eq.~({\ref{hq}), it is seen that the hybrid ensemble Hamiltonian comprises (i) a quantum-like term corresponding to free centre-of-mass motion but with a rescaled Planck constant 
\[ \hbar_{\overline{X}} := [m/(m+M)]^{1/2}\, \hbar ; \]
(ii) a quantum-like term corresponding to relative motion in a potential $V(r)$ but with a rescaled Planck constant
\[ \hbar_R := [M/(m+M)]^{1/2}\, \hbar ; \]
and (iii) an intrinsic interaction term \cite{scat}.  The presence of this last term implies, in contrast to classical-classical and quantum-quantum interactions, that {\it the centre-of-mass motion and the relative motion do not decouple for quantum-classical interactions}.

\section{Measurement and decoherence}

In the standard Copenhagen interpretation of quantum mechanics, it has been repeatedly emphasised that any objective account of a physical experiment must be given in classical terms.  Thus, for example, Bohr stated that {\it ``the point is that in each case we must be able to communicate to others what we have done and what we have learned, and that therefore the functioning of the measuring instruments must be described within the framework of classical physical ideas"} (page 89 of Ref.~\cite{bohr}), while Heisenberg wrote that {\it ``the concepts of classical physics form the language by which we describe the arrangements of our experiments and state the results"} 
(page 46 of Ref.~\cite{heis}; see also page 127 onwards for an extended discussion).  
Hence, if the Copenhagen interpretation is to be taken seriously, it follows that any dynamical description of the measurement process should be able to be formulated, at least approximately, in terms of an interaction between classical and quantum systems.  It is of obvious interest to determine whether this can be done using the formalism of Sec.~II above.

Consider first the case of a measurement of position on members of an ensemble of quantum particles of mass $m$.  Now, the fully quantum description of an interaction which linearly couples the quantum position $q$ to a quantum pointer position $x$ is described by the Hamiltonian operator $\kappa(t)\hat{p}_C\,\hat{q}$, where $\hat{p}_C$ denotes the operator conjugate to the pointer observable.  Following the prescription given in Sec.~II.B, the composite ensemble Hamiltonian describing the corresponding interaction between an ensemble of quantum particles and an ensemble of {\it classical} pointers is then found to have the form
\begin{equation} \label{qmeas}
\tilde{H}_{\rm position} = \tilde{H}_{QC} + \kappa(t) \int dq\,dx\, P\, q.\nabla_x S ,
\end{equation}
where $\tilde{H}_{QC}$ is given in Eq.~(\ref{vqx}).  It is worth noting that the interaction term would in fact have precisely the same form if the two ensembles were both quantum or both classical, and hence this `position measurement' interaction is of a universal type.

It is convenient to assume that the measurement interaction occurs over a sufficiently short time period, $[0,T]$, such that $\tilde{H}_{QC}$ can be ignored during the interaction.  The equations of motion then follow via Eqs.~(\ref{motioncont}) and (\ref{qmeas}) as
\[ \frac{\partial P}{\partial t} = -\kappa(t)\, q.\nabla_x P ~~~~~~ \frac{\partial S}{\partial t} = -\kappa(t)\, q.\nabla_x S , \]
which may be trivially integrated over the interaction period to give
\begin{equation} \label{qsol}
P(q,x,T) = P(q, x-Kq, 0),~~~~~~S(q,x,T) = S(q, x-Kq,0) ,
\end{equation}
where $K:=\int_0^T dt\,\kappa(t)$. Thus, as expected, the interaction directly correlates the classical pointer position $x$ with the quantum particle position $q$.  

In particular, if the initial position of the pointer is sharply defined as some value $x_0$ for each member of the ensemble, so that $P(q,x,0) = \delta(x-x_0)P_Q(q)$, then Eq.~(\ref{qsol}) implies a spread of pointer positions over the ensemble after the measurement, with result $x_0-Kq$ obtained with probability $P_Q(q)$.  Moreover, the conditional density operator after measurement follows via Eqs.~(\ref{cond}) and (\ref{rho}) as
\[ \rho_{Q|C} = \int dq \,P_Q(q)\, |q\rangle\langle q| ,\]
and thus `decoheres' with respect to position, i.e., it becomes diagonal in the position basis.

For an example involving the measurement of spin, consider the composite ensemble Hamiltonian $\tilde{H}_{\rm spin}$ in Eq.~(\ref{hspin}), which may be interpreted as coupling an ensemble of  
one-dimensional classical pointers to the $z$-component of an ensemble of quantum spin-1/2 particles.  Assuming as before that the measurement interaction occurs over a sufficiently short time period $[0,T]$, the first term of $\tilde{H}_{\rm spin}$ can be ignored, and the equations of motion during the interaction follow via 
Eq.~(\ref{motioncont}) as
\[ \frac{\partial P(x,\alpha,t)}{\partial t} = -\alpha\,\kappa(t)\frac{dP(x,\alpha)}{dx},~~~~~~\frac{\partial S(x,\alpha,t)}{\partial t} = -\alpha\,\kappa(t)\frac{dS(x,\alpha)}{dx} .\]
These can be trivially integrated to give
\begin{equation} \label{spinsol}
P(x,\alpha,T) = P(x-\alpha K,\alpha, 0),~~~~~~~~S(x,\alpha, T) = S(x-\alpha K,\alpha, 0),
\end{equation}
where again $K:=\int_0^T dt\,\kappa(t)$.  Thus, as expected, the interaction directly correlates the pointer position $x$ with the spin $\alpha$ in the $z$-direction.

In particular, if the classical and quantum ensembles are initially independent (see Sec.~II.B), with the classical ensemble described by the conjugate quantities $P_C(x)$ and $S_C(x)$ and the quantum ensemble by the spinor with components $\psi_\pm=(w_\pm)^{1/2}e^{i\sigma_\pm /\hbar}$ in the $\sigma_z$-basis, then
\[ P(x,\pm,0) = w_\pm \,P_C(x),~~~~~~S(x,\pm,0) = S_C(x) + \sigma_\pm ,\]
and the marginal probability density for the pointer position after measurement follows via Eq.~(\ref{spinsol}) as the mixture
\[ P(x,T) = \sum_{\alpha=\pm} P(x,\alpha,T) = w_+\, P_C(x-K) + w_- \,P_C(x+K) .\]
Hence, the initial probability density $P_C(x)$ is displaced by $K$ with probability $w_+$, and by $-K$ with probability $w_-$, where $w_\pm$ denotes the initial probability of spin up/down in the $z$-direction.  Further, if the initial pointer probability density $P_C(x)$ has a spread which is small with respect to $K$, so that $P_C(x-K)\, P_C(x+K)\approx 0$, then the conditional density matrix after the measurement follows via Eqs.~(\ref{cond}) and (\ref{rho}) as
\[ \rho_{Q|C} = w_+\, |+\rangle\langle +| + w_-\, |-\rangle\langle -| ,\]
and hence `decoheres' with respect to the $\sigma_z$ basis.  It may be shown that this result is in fact independent of the basis used to represent the quantum component of the composite ensemble, i.e., decoherence with respect to the $\sigma_z$ basis is a consequence of the interaction, not of the basis chosen to represent the quantum component of the ensemble.

The above examples demonstrate the consistency of describing measurements via classical-quantum interactions, as required by the Copenhagen interpretation.  In particular, (i) the measuring apparatus is described classically, as is required for the unambiguous communication and comparison of physical results; (ii) information about quantum ensembles is obtained via an appropriate interaction with an ensemble of classical measuring apparatuses, which correlates the classical configuration with a corresponding quantum property, and (iii) there is a conditional decoherence of the quantum ensemble relative to the classical ensemble, which depends upon the nature of the quantum-classical interaction.  

Finally, note that the question of `where' to place the quantum-classical cut is operationally trivial in this approach: any objective description of an experimental setup and its results must be a classical description, and so the cut may be placed at that point.  This is in direct contrast to the measurement problem that arises in approaches that attempt to describe the measuring apparatus as a quantum object.

\section{Hybrid oscillators and coherent states}

The hybrid harmonic oscillator is, like its fully quantum and fully classical counterparts, a relatively simple system to investigate.  Here it is shown that the notion of `coherent states' is straightforward to generalise to the hybrid case.

Consider a composite ensemble comprising $n$-dimensional quantum particles of mass $m$ joined by springs to $n$-dimensional classical particles of mass $M$.  The corresponding composite ensemble Hamiltonian then has the form of $\tilde{H}_{QC}$ in Eq.~(\ref{vqx}), with interaction potential
\begin{equation} \label{spring} 
V(q,x,t) = \frac{1}{2}k|q-x|^2 .  
\end{equation}
It is convenient to define hybrid coordinates $\xi:=(q,x)$, and block matrices
\begin{equation} \label{block}
C:= \left(
\begin{array}{cc}
kI & -kI\\-kI & kI
\end{array} 
\right),~~~
U:= \left(
\begin{array}{cc}
m^{-1}I & 0\\0 & M^{-1}I
\end{array} 
\right),~~~
E:= \left(
\begin{array}{cc}
I & 0\\0 & 0
\end{array} 
\right)
\end{equation}
(where $I$ denotes the $n\times n$ identity matrix), allowing 
$\tilde{H}_{QC}$ to be rewritten as
\begin{equation} \label{hosc} 
\tilde{H}_{\rm osc}= \int d\xi\, P \left[ \frac{1}{2}(\nabla S)^T U\nabla S + \frac{1}{2}\xi^T C\xi + \frac{\hbar^2}{8} (\nabla \log P)^T EUE (\nabla \log P) \right] .
\end{equation}
Note that the classical-classical oscillator corresponds to replacing  $E$ by $0$, and that the quantum-quantum oscillator corresponds to replacing $E$ by $1$.  This is useful for comparisons between the three cases.

The equations of motion follow from Eqs.~(\ref{motioncont}) and 
(\ref{hosc}) as
\begin{equation} \label{posc}
\frac{\partial P}{\partial t} + \nabla .\left(PU\nabla S\right) = 0 ,\end{equation}
\begin{equation} \label{sosc}
\frac{\partial S}{\partial t} + \frac{1}{2}(\nabla S)^T U\nabla S + \frac{1}{2}\xi^T C\xi -\frac{\hbar^2}{2}\frac{\nabla .(EUE\nabla 
\sqrt{P})}{\sqrt{P}} =0 . 
\end{equation}
Recalling that quantum coherent states have Gaussian probability densities, it is natural to look for solutions of the form
\begin{equation} \label{pgauss} 
P(\xi,t) = \frac{\sqrt{\det K}}{(2\pi)^{n}} e^{-\frac{1}{2} (\xi -\alpha)^T K(\xi -\alpha)} 
\end{equation}
for some (possibly time-dependent) positive definite symmetric matrix $K$ and vector $\alpha$.  
This  Gaussian ansatz is consistent with the equations of motion if (and only if) $S$ is at most quadratic in $\xi$.  For simplicity, it will in fact be further assumed that $S$ has the {\it linear} form
\begin{equation} \label{slin} 
S(\xi,t) = \beta .(\xi-\alpha) + \sigma  , 
\end{equation}
where $\beta$ is a vector and $\sigma$ is a scalar (both possibly 
time-dependent).

The equations of motion for $K$, $\alpha$, $\beta$ and $\sigma$ may be found by substituting the above forms of $P$ and $S$ into 
Eqs.~(\ref{posc}) and (\ref{sosc}), and equating coefficients of the respective quadratic, linear and constant terms with respect to $\xi-\alpha$.  After some straightforward algebra (requiring the formula
$(d/dt)\det K = \det K~{\rm tr}[\dot{K}K^{-1}]$), one obtains 
\begin{equation} \label{kab}
\dot{\alpha} = U\beta;~~~~~\dot{\beta}=-C\alpha,~~~~~
\dot{K}=0,~~~~~KEUEK= \frac{4}{\hbar^2}C,
\end{equation}
\begin{equation} \label{sigma}
\dot{\sigma} = \frac{1}{2}\dot{(\alpha.\beta)} - \frac{\hbar^2}{4} {\rm tr}[EUEK] = \frac{1}{2}\dot{(\alpha.\beta)} -{\rm tr}[CK^{-1}].
\end{equation}
Note that the first three equations and the last equation are independent of the projection matrix $E$, and hence are also valid for classical-classical and quantum-quantum oscillators.

The above equations for $\dot{\alpha}$ and $\dot{\beta}$ are precisely those corresponding to a classical-classical oscillator described by configuration $\alpha$, conjugate momentum $\beta$, and Hamiltonian function
\[  H(\alpha,\beta) = \frac{1}{2}\beta^TU\beta + \frac{1}{2}\alpha^T C\alpha .  \]
Thus, solving for $\alpha$ and $\beta$ is equivalent to solving the classical equations of motion.  Noting Eq.~(\ref{pgauss}), {\it the probability density is therefore a Gaussian centred on the classical motion}.  This link to classical motion is, of course, a desirable feature of `coherent states'.
Using the forms of $C$ and $U$ in Eq.~(\ref{block}), the general solutions for $\alpha$ and $\beta$ are
\begin{eqnarray} \label{asol}
\alpha &=& (c,c) + \mu \cos \left( 
\omega_\mu t+\phi\right)~(d/m,-d/M) ,\\ \label{bsol}
\beta &=& -\mu \omega_\mu \sin \left( \omega_\mu t +\phi\right)~(d,-d) ,
\end{eqnarray}
as may be checked by direct substitution, where $c$ and $d$ are arbitrary $n$-vectors, $\phi$ is an arbitary constant, $\mu$ is the reduced mass defined in Eq.~(\ref{mass}), and $\omega_{\mu}:=\sqrt{k/\mu}$.  It is noteworthy that the frequency $\omega_{\mu}$ associated with the motion is determined by the reduced mass, given that the centre-of-mass and relative motions do not decouple (see Sec.~II.C).

Further, the first equation for $K$ in Eq.~(\ref{kab}) implies that it is constant, while the second can be solved for $K$ by writing it out in block-matrix form, yielding 
\begin{equation} \label{ksol}  K=\frac{2}{\hbar}\sqrt{\frac{m}{k}} C + 
\left( \begin{array}{cc} 0&0\\0&A\end{array} \right) , 
\end{equation}
where $A$ is any nonnegative symmetric $n\times n$ matrix.  Substituting 
Eqs.~(\ref{asol}) and (\ref{ksol}) into Eq.~(\ref{pgauss}) then yields the general solution
\[
P(q,x,t) = P_A(x-x_t) (\sqrt{km}/\pi\hbar)^{n/2} e^{-\sqrt{km}|q-x-(q_t-x_t)|^2/\hbar} 
\]
for the probability density, where 
\[ q_t = c + (\mu/m)d \cos \left( \omega_\mu t+\phi\right),~~~~~
x_t = c -(\mu/M)d \cos \left( \omega_\mu t+\phi\right) \]
denote the quantum and classical components of $\alpha$, and $P_A(x)$ denotes the Gaussian probability density 
\[  P_A(x) := (2\pi)^{-n/2} (\det A)^{1/2}e^{-\frac{1}{2} x^TAx} . \]

Clearly, the dispersion of the above probability density is minimised in the limit that $P_A(x)$ approaches a delta-function (i.e., $A^{-1}\rightarrow 0$), and hence {\it this limit will be taken to define the `coherent state' solutions for the hybrid oscillator}.  The corresponding probability density follows as
\begin{equation} \label{psol}
P(q,x,t)= \delta(x-x_t) (\sqrt{km}/\pi\hbar)^{n/2} e^{-\sqrt{km}|q-q_t|^2/\hbar} ,
\end{equation}
i.e., {\it for hybrid coherent states, the position of each classical particle is described by the trajectory $x_t$, while the positions of the quantum particles are described by a Gaussian of width $\hbar/2\sqrt{km}$ in each direction, centred on the trajectory $q_t$}.

Finally, substituting Eqs.~(\ref{asol}), (\ref{bsol}) and (\ref{ksol}) into Eqs.~(\ref{slin}) and (\ref{sigma}) yields 
\begin{equation} \label{ssol}
S(q,x,t) = -\frac{1}{2}n\hbar\omega_m t + \frac{|d|^2}{4}\sqrt{k\mu} \sin 2(\omega_\mu t+ \phi) - d.(q-x)\sqrt{k\mu} \sin (\omega_\mu t+ \phi) 
\end{equation}
for $S$, up to an arbitary additive constant, where $\omega_m:=\sqrt{k/m}$.  Thus there are {\it two} natural frequencies, $\omega_\mu$ and $\omega_m$, associated with hybrid coherent states.
Note that the choice $d=0$ corresponds to a `stationary state' as defined in Sec.~II.C.  In this case only the first term of $S(q,x,t)$ above is non-vanishing, and the numerical value of the composite ensemble Hamiltonian may be calculated from Eq.~(\ref{stat}) as
\[ \tilde{H}_{\rm osc} = -\frac{\partial S}{\partial t} =\frac{1}{2}n\hbar\omega_m ,\]
which may be recognised as the zero-point energy of an n-dimensional quantum harmonic oscillator of mass $m$, as expected.

\section{Coupling quantum fields to classical spacetime}

Although much effort has been devoted to the problem of quantizing gravity,
there are a number of conceptual and technical issues that still remain
unsolved. Some of these difficulties have been turned into arguments \textit{%
against} the quantization of gravity, and it has been suggested that
spacetime should be treated instead as a classical field that is in no need
of quantization \cite{Rosenfeld}. For example, Freeman Dyson \cite{Dyson}
has recently suggested that it might be impossible in principle to observe
the existence of individual gravitons, and this has lead him to conclude
that "the gravitational field described by Einstein's theory of general
relativity is a purely classical field without any quantum
behaviour\textquotedblright . Isham and Butterfield \cite{Butterfield},
while putting forward the point of view that some type of theory of quantum
gravity should be sought, have concluded that there is arguably no
definitive proof that general relativity \textit{has} to be quantized. The
literature on this topic is extensive. Some recent discussions of arguments
against a quantum theory of gravity and counter-arguments in favour of such
a theory, as well as critiques of the semi-classical approach mentioned
below, can be found in Refs.~\cite{kiefer,carlip,Isham}.

If gravity ought to remain classical in nature, then the crucial question is
perhaps not how to quantize gravity but rather how to couple quantum fields
to a classical spacetime (this is of course also important as long as a
quantum theory remains elusive). The standard way of doing this leads to the
semiclassical Einstein equations, in which the energy mometum tensor that
serves as the source in the Einstein equations is replaced by the
expectation value of the energy momentum operator $\widehat{T}_{\mu \nu }$
with respect to some quantum state $\Psi $:%
\begin{equation}
^{4}R_{\mu \nu }-\frac{1}{2}g_{\mu \nu }\,^{4}R+\Lambda g_{\mu \nu }=\frac{%
\kappa }{2}\,\left\langle \Psi \right\vert \widehat{T}_{\mu \nu }\left\vert
\Psi \right\rangle   \label{SCEE}
\end{equation}%
where $^{4}R_{\mu \nu }$ is the curvature tensor, $^{4}R$\ the curvature
scalar and $g_{\mu \nu }$ the metric tensor in spacetime, $\Lambda $ is the
cosmological constant and $\kappa =16\pi G$ (we use units where $c=1$). But
this approach presents a number of well known difficulties, and it is
therefore not suitable if we want to model the classical-quantum interaction
in a realistic way. The formalism of configuration-space ensembles provides
an alternative to Eq.~(\ref{SCEE}).

Before discussing this alternative, we summarize some results concerning the
ensemble formalism for pure gravity \cite{hkr,HGRG,RBJP}. The most direct
way of introducing a classical configuration-space ensemble is to start from
the Hamilton-Jacobi equation, which in the metric representation takes the
form 
\begin{equation} \label{hg}
\mathcal{H}_{h}=\kappa G_{ijkl}\frac{\delta S}{\delta h_{ij}}\frac{\delta S}{%
\delta h_{kl}}-\frac{1}{\kappa }\sqrt{h}\left( R-2\Lambda \right) =0,
\end{equation}%
where $R$\ is the curvature scalar and $h_{kl}$ the metric tensor on a
three-dimensional spatial hypersurface, $G_{ijkl}=\left( 2h\right)
^{-1/2}\left( h_{ik}h_{jl}+h_{il}h_{jk}-h_{ij}h_{kl}\right) $ is the DeWitt
supermetric, and the functional $S$ is assumed to be invariant under the
gauge group of spatial coordinate transformations \cite{MTW}. An appropriate
ensemble Hamiltonian is given by 
\[
\tilde{H}_{h}=\int d^{3}x\int Dh\,P\,\mathcal{H}_{h}  \]
(technical issues are discussed in more detail in \cite{RBJP}). The
equations of motion then have the form%
\[
\frac{\partial P}{\partial t}=\frac{\Delta \tilde{H}_{h}}{\Delta S},\quad 
\frac{\partial S}{\partial t}=-\frac{\Delta \tilde{H}_{h}}{\Delta P}
\]%
where $\Delta /\Delta F$ denotes the variational derivative with respect to
the functional $F$ \cite{hkr}. Assuming the constraints $\frac{\partial S}{\partial t}=$
$\frac{\partial P}{\partial t}=0$, these equations of motion lead to Eq.~(\ref%
{hg}) and the continuity equation 
\[
\int d^{3}x\frac{\delta }{\delta h_{ij}}\left( PG_{ijkl}\frac{\delta S}{%
\delta h_{kl}}\right) =0.
\]

Consider now the case where a quantum scalar field $\phi$ couples to
the classical metric $h_{kl}$. Then, $\tilde{H}_{h}$ has to be
generalized according to the method of Sec.~II, to
\begin{equation} \label{genphi}
\tilde{H}_{\phi h}=\int d^{3}x\int Dh\,P\left[ \mathcal{H}_{\phi h}+Q_{\phi }%
\right] , 
\end{equation}%
where (cf. Eq.~(4.74) of Ref.~\cite{kiefer})%
\[
\mathcal{H}_{\phi h}=\mathcal{H}_{h}+\frac{1}{2\sqrt{h}}\left( \frac{\delta S%
}{\delta \phi }\right) ^{2}+\sqrt{h}\left[ \frac{1}{2}h^{ij}\frac{\partial
\phi }{\partial x^{i}}\frac{\partial \phi }{\partial x^{j}}+V\left( \phi
\right) \right] , 
\]%
and where%
\[
Q_{\phi }=\frac{\hbar ^{2}}{4}\frac{1}{2\sqrt{h}}\left( \frac{\delta \log P%
}{\delta \phi }\right) ^{2} 
\]%
is the additional, non-classical kinetic energy term.

As a simple example of a quantum ensemble of fields interacting with a
classical ensemble of metrics, we consider the case of a closed
Robertson-Walker 
universe with a massive scalar field in the minisuperspace model \cite%
{kiefer}. The line element is assumed to be of the form%
\[
ds^{2}=-N^{2}\left( t\right) dt^{2}+a^{2}\left( t\right) d\Omega _{3}^{2}
\]%
where $a$ is the scale factor and $d\Omega _{3}^{2}$ is the standard line
element on $S^{3}$. After symmetry reduction, the problem admits a
minisuperspace formulation in a finite dimensional configuration space with
coordinates $\left\{ a,\phi \right\} $. For simplicity, we restrict to a
potential term that is quadratic in $\phi $. The classical Hamilton-Jacobi
equation takes the form $\mathcal{H}_{\phi a}=0$, with%
\[
\mathcal{H}_{\phi a}:=-\frac{1}{a}\left( \frac{\partial S}{\partial a}\right)
^{2}+\frac{1}{a^{3}}\left( \frac{\partial S}{\partial \phi }\right) ^{2}-a+%
\frac{\Lambda a^{3}}{3}+m^{2}a^{3}\phi ^{2} ,
\]%
where $m$ is the mass of the field and we use units for which $2G/3\pi =1$.
The corresponding ensemble Hamiltonian for a quantum field interacting with
the classical metric is given in this model by%
\[
\tilde{H}_{\phi a}=\int dad\phi P\left[ \mathcal{H}_{\phi a}+\frac{\hbar ^{2}%
}{4}\frac{1}{a^{3}}\left( \frac{\partial \log P}{\partial \phi }\right) ^{2}%
\right] .
\]%
Assuming once more the constraints $\frac{\partial S}{\partial t}=$ $\frac{%
\partial P}{\partial t}=0$, the equations of motion follow as%
\begin{equation}
\mathcal{H}_{\phi a}-\frac{\hbar ^{2}}{a^{3}}\frac{1}{\sqrt{P}}\frac{%
\partial ^{2}\sqrt{P}}{\partial \phi ^{2}}=0  \label{HQ}
\end{equation}%
and%
\begin{equation}
-\frac{\partial }{\partial a}\left( \frac{P}{a}\frac{\partial S}{\partial a}%
\right) +\frac{\partial }{\partial \phi }\left( \frac{P}{a^{3}}\frac{%
\partial S}{\partial \phi }\right) =0.  \label{C}
\end{equation}

An \textit{exact solution} can be derived provided we introduce the ansatz $%
S=0$. Then, Eq.~(\ref{HQ}) reduces to%
\begin{equation}
-\frac{\hbar ^{2}}{a^{3}}\frac{1}{\sqrt{P}}\frac{\partial ^{2}\sqrt{P}}{%
\partial \phi ^{2}}-a+\frac{\Lambda a^{3}}{3}+a^{3}m^{2}\phi ^{2}=0
\label{HHJ}
\end{equation}%
while Eq.~(\ref{C}) is automatically satisfied. The non-negative,
normalizable solutions of Eq.~(\ref{HHJ}) take the form%
\begin{equation}
P_{n}\left( \phi ,a\right) =\delta \left( a-a_{n}\right) \frac{\lambda _{n}}{%
\sqrt{\pi }2^{n}n!}\exp \left( -\lambda _{n}^{2}\phi ^{2}\right) \,
\left[H_{n}\left( \lambda _{n}\phi \right)\right]^2   \nonumber
\end{equation}%
where the $H_{n}$ are Hermite polynomials, $\lambda
_{n}^{2}=a_{n}^{3}m/\hbar \ $and the $a_{n}$\ satisfy the condition%
\begin{equation} \label{quanta}
a_{n}-\frac{\Lambda a_{n}^{3}}{3}=2\hbar m\left( n+\frac{1}{2}\right) 
\end{equation}%
for $n=\left\{ 0,1,2,...\right\} $. It is remarkable that {\it the coupling of
the quantum field to a purely classical metric leads to a
quantization condition for the scale factor $a$}. Further, the
classical singularity at $a=0$ is \textit{excluded} from these solutions. If
the term proportional to the cosmological constant $\Lambda $ can be
neglected, the quantization condition takes the simple form $a_{n}=2\hbar
m\left( n+\frac{1}{2}\right) $. It is possible to define $\psi =\sqrt{P}$
and derive a Schr\"{o}dinger equation for $\psi $ from Eq. (\ref{HHJ}).
However, it is not possible to introduce a linear superposition of the $\psi
_{n}$ because the potential term in this equation ends up being a function
of $a_{n}$.  The quantization condition Eq.~(\ref{quanta}) can also be obtained by replacing the ansatz $S=0$ with the constraint that the radius is well-defined, i.e., $P(\phi,a)=\delta(a-r)\,P_0(\phi)$ for some fixed value $r$.

We close this section with some remarks concerning solutions for the case of
potentials that include other $\phi $-dependent terms in addition to the
term quadratic in $\phi $ that we have considered in our minisuperspace
example. In all cases, the ansatz $S=0$ [or, alternatively, $P(\phi,a)=\delta(a-r)P_0(\phi)$], will lead to equations that reduce
to the form of a time-independent Schr\"{o}dinger equation, with
a modified potential term that is a function of $a$, and an energy term given
by $E=a-\Lambda a^{3}/3$. If the solution of this Schr\"{o}dinger equation
only admits discrete energy levels $E_{n}$, we will be lead again to a
quantization condition for the scale factor, of the form $E_{n}=a_{n}-\Lambda
a_{n}^{3}/3$. Thus, the quantization of the scale factor is a generic
feature of such models. In particular, consider the case in which the
modified potential remains non-negative and $\Lambda=0$. 
Then, the ground state energy $E_{n}
$ is strictly positive, and the minimum value of the scale factor is given
by $a_{0}=E_{0}$. Hence, the quantum fluctuations associated with the
matter fields in the ground state may be interpreted as being directly
responsible for removing the classical singularity. 

Finally, we would like
to point out that the solutions $\left\{ P_{n}\right\} $ that correspond to
a given modified potential have an interesting property that might be of
relevance to a resolution of the problem of the arrow of time. While there
is no external time parameter in quantum cosmology, one may introduce an
intrinsic time parameter defined in terms of the radius $a$ (or any increasing function of 
$a$ such as $\log a$). It has previously been argued by Zeh and Kiefer that, with appropriate initial conditions, the solutions of
the Wheeler-DeWitt equation for a Robertson-Walker 
model with small perturbations
will have the property that the entropy, suitably defined, increases with
increasing scale factor (this is connected to the asymmetry of the potential
term with respect to the intrinsic time, and in particular to the property that the
potential vanishes as $a\rightarrow 0$) \cite{kiefer,zehkief}. 
In a similar way, there is a natural ordering of the solutions $\left\{
P_{n}\right\}$, in terms of a discrete time variable given by $n$, and
this
ordering leads to a thermodynamic arrow of time.  This follows from the
observation that the amount of structure associated with a solution
$P_{n}$
(as determined, for example, by counting the number of nodes in $\psi
_{n}$
or by evaluating the entropy expression 
$-\int d\phi\, P_{n}\log P_{n}$ for
different values of $n$)\ increases with increasing $n$. 
Note that this
thermodynamic arrow of time coincides with the arrow of time as
determined
by an expanding universe whenever the non-linear term proportional to
$\Lambda$ can be neglected.

\section{Conclusions}

It has been shown that the formalism of configuration-space ensembles allows a general and consistent description of interactions between quantum and classical ensembles.  Moreover, this description has been successfully applied to examples as diverse as measurement, hybrid oscillators, and the interaction of quantum matter with classical spacetime.

Each of the above examples suggest topics for future examination.  For example, while most physicists would agree that the question ``What is measurement?" is closely related to the question ``What is classical?", many would also agree that the Copenhagen interpretation does not provide a clear answer to either question.  Hence various extra elements have been suggested, including decoherence, spontaneous localisation, nonlocal guiding waves, branching universes, and even human consciousness.  The results in Secs.~II and III suggest that it is the physical appropriateness of describing the dynamics of a given system by a {\it classical} ensemble Hamiltonian that provides the basis for defining classical behaviour more generally.
Note also that, in the Copenhagen interpretation, the unavoidable restriction to the language of classical physics for describing measurements necessarily leads, via the existence of physically incompatible or `complementary' experimental arrangements, to fundamental limits on the description of quantum phenomena. These limits are typically quantified by statistical uncertainty relations, and it would be of some interest to investigate how such relations might emerge in the above formalism.

It would similarly be of interest to extend the results in Secs.~II.C, IV and V, including studying the scattering of quantum particles from classical particles; finding more general solutions to the equations of motion in Eqs.~(\ref{posc}) and (\ref{sosc}) for the hybrid harmonic oscillator; and applying the equations of Sec.~IV to problems in black hole thermodynamics and more general problems of cosmology.

More generally, the configuration ensemble formalism itself requires further investigation.  One important question is to what extent the theory of unitary transformations for quantum ensembles, relating complementary configuration spaces, can be generalised.  For example, while the mapping $P\rightarrow P+S$, $S\rightarrow (S-P)/2$ is formally a canonical transformation of the canonically conjugate quantities $P$ and $S$, neither of the new quantities can be interpreted as a conserved probability density on some configuration space.  It is straightforward to show that the set of `physical' canonical transformations, which preserve the normalisation and positivity of $P$, are generated by functionals $G[P,S]$ satisfying 
Eq.~(\ref{invar}) with $\tilde{H}$ replaced by $G$ \cite{gen}.  For continuous configuration spaces such generators include the `ensemble momentum' $\Pi=\int dx\, P\nabla S=\langle \nabla S\rangle$ \cite{super}, and its canonical conjugate $X=\int dx\, Px=\langle x\rangle$, which satisfy $\{ X_j,P_k\}=\delta_{jk}$ and which generate translations of $x$ and $\nabla S$ respectively \cite{ang}:
\[ a.\Pi : P(x),S(x)\rightarrow P(x-a),S(x-a),~~b.X : P(x),S(x)\rightarrow P(x),S(x)-b.x~.\]
These transformations are important in establishing the relationship between the position and momentum representations in quantum mechanics \cite{hrjpa}, and are expected to have a similar role more generally.


\begin{thebibliography}{99}

\bibitem{bohr} N. Bohr, {\it Atomic Physics and Human Knowledge} (Wiley, New York, 1958).
\bibitem{heis} W. Heisenberg, {\it Physics and Philosphy} (Allen and Unwin, London, 1958), Chaps.~3, 8.
\bibitem{boucher} W. Boucher and J. Traschen, Phys. Rev. D {\bf 37}, 3522 (1988).
\bibitem{meanfield} N. Makri, Annu. Rev. Phys. Chem. {\bf 50}, 167 (1999)
\bibitem{hawking} S.W. Hawking, Commun. Math. Phys. {\bf 43}, 199 (1975).
\bibitem{kiefer} C. Kiefer, {\it Quantum Gravity} (Clarendon Press, Oxford, 2004), Secs.~1.2, 5.4, 8.1, 10.2.
\bibitem{diosihalli} L. Diosi and J.J. Halliwell, Phys. Rev. Lett. {\bf 81}, 2846 (1998).
\bibitem{hu} B.L. Hu and E. Verdaguer, Class. Quantum Grav. {\bf 20}, R1 (2002); eprint gr-qc/0211090.
\bibitem{sudpramana} E.C.G. Sudarshan, Pramana {\bf 6}, 117 (1976).
\bibitem{sud1} T.N. Sherry and E.C.G. Sudarshan, Phys. Rev. D {\bf 18}, 4580 (1978).
\bibitem{sud2} T.N. Sherry and E.C.G. Sudarshan, Phys. Rev. D {\bf 20}, 
857 (1979)
\bibitem{sud3} S.R. Gautam, T.N. Sherry, and E.C.G. Sudarshan, Phys. Rev. D {\bf 20}, 3081 (1979).
\bibitem{salcedo1} L.L. Salcedo, Phys. Rev. A {\bf 54}, 3657 (1996).
\bibitem{salcedo2} J. Caro and L.L. Salcedo, Phys. Rev. A {\bf 60} 842 (1999).
\bibitem{royal} L. Diosi, N. Gisin, and W.T. Strunz, Phys. Rev. A {\bf 61}, 022108 (2000).
\bibitem{ternoperes} A. Peres and D.R. Terno, Phys. Rev. A {\bf 63}, 022101 (2001).
\bibitem{sahoo} D. Sahoo, J. Phys. A {\bf 37}, 997 (2004).
\bibitem{dias} N.C. Dias and J.N. Prata, eprint quant-ph/0005019.
\bibitem{bohm} P.R. Holland, {\it The Quantum 
Theory of Motion} (Cambridge University Press, Cambridge, 1993), Chap. 3.
\bibitem{traj} E. Gindensperger, C. Meier, and J.A. Beswick, J. Chem. Phys. {\bf 113} 9369 (2000); O.V. Prezhdo and C. Brooksby, Phys. Rev. Lett. {\bf 86}, 3215 (2001); I. Burghardt and G. Parlant, J. Chem. Phys. {\bf 120}, 3055 (2004).
\bibitem{dbbprobs} O. Passon, eprint quant-ph/0412119.  
\bibitem{salcedocom} L.L. Salcedo, Phys. Rev. Lett. {\bf 90}, 118901 (2003).
\bibitem{prezreply} O. Prezhdo and C. Brooksby, Phys. Rev. Lett. {\bf 90}, 118902 (2003).
\bibitem{super} M.J.W. Hall, J. Phys. A {\bf 37}, 7799 (2004).
\bibitem{goldstein} H. Goldstein, {\it Classical Mechanics} 
(Addison-Wesley, New York, 1950), Chaps.~7, 8, 11.
\bibitem{constraint} A further difference is the topological constraint imposed on quantum ensembles of particles, namely that $\oint_C dS/h$ is an integer for all loops $C$ in configuration space \cite{bohm} (corresponding to single-valuedness of the wave function $\psi=P^{1/2}e^{iS/\hbar}$). It is important to note that a similar constraint may also be applied to purely classical ensembles, as was done for example by R. Schiller, Phys. Rev. \textbf{125}, 1100 (1962),
when he formulated his theory of `quasi-classical mechanics', which deals with a particular class of classical ensembles (those that can be associated with WKB solutions). Thus, constraints of this type are consistent with the equations of motion derived from either $\tilde{H}_C$ or $\tilde{H}_Q$, amounting only to a reduction of the space of physically allowed solutions, and are logically independent of the choice of the ensemble Hamiltonian.  For a general theory of constraints in the configuration ensemble formalism, see Ref.~\cite{super}.
\bibitem{frieden} B.R. Frieden, Am. J. Phys. {\bf 57}, 1004 (1989).
\bibitem{regpra} M. Reginatto, Phys. Rev. A {\bf 58}, 1775 (1998).
\bibitem{hrjpa} M.J.W. Hall and M. Reginatto, J. Phys. A {\bf 35}, 3289 (2002).
\bibitem{hkr} M.J.W. Hall, K. Kumar and M. Reginatto, J. Phys. A {\bf 36}, 9779 (2003).
\bibitem{roman} P. Roman, {\it Advanced Quantum Theory} (Addison-Wesley, Reading, 1965), pp. 31-33. 
\bibitem{nik} H. Nikolic, eprint quant-ph/0505143.
\bibitem{interact} Note that not all pairs of ensemble Hamiltonians $\tilde{H}_A$ and $\tilde{H}_B$ are compatible with the existence of a composite ensemble Hamiltonian $\tilde{H}_{AB}$ satisfying Eq.~(\ref{nonint}).  In such cases the corresponding ensembles can physically co-exist only if there is an ongoing irreducible interaction between them - an interaction that cannot be `switched off', even in principle. It is conjectured that a necessary (but not sufficient) condition, for the extendibility of two ensemble Hamiltonians $\tilde{H}_A$ and $\tilde{H}_B$ to a non-interacting composite ensemble Hamiltonian, is that they are homogenous of degree unity with respect to the probability density.  This condition also implies that the quantity $-P\partial S/\partial t$ may be interpreted as a local energy density on configuration space, and is satisfied for all quantum ensembles \cite{super}.  
\bibitem{con} In the case that the quantum component has a continuous configuration space, it is natural to also impose the topological constraint that $\oint_{C_x} dS/h$ is an integer for all loops $C_x$ in the joint configuration space that correspond to some fixed classical configuration $x$, corresponding to singe-valuedness of the conditional wavefunction in Eq.~(\ref{cond}) (see also footnote~\cite{constraint} above).
\bibitem{scat} The only limits in which this interaction term can be ignored are (i) the limit of an infinite classical or quantum mass ($M\rightarrow\infty$ or $m\rightarrow\infty$), for which the ensemble Hamiltonian reduces to the sum of a (classical or quantum) centre-of-mass term and a (quantum or classical) relative motion term; and (ii) the limit of a vanishing Planck constant, $\hbar\rightarrow 0$, for which the ensemble Hamiltonian reduces to the sum of two classical terms.  
\bibitem{Rosenfeld} L. Rosenfeld, Nucl. Phys. \textbf{40}, 353 (1963).

\bibitem{Dyson} F. Dyson, The New York Times Book Review, 2004/05/28.

\bibitem{Butterfield} J. Butterfield and C. Isham, Spacetime and the
philosophical challenge of quantum gravity. In: \textit{Physics meets
philosophy at the Planck scale}, C. Callender and N. Huggett, eds.
(Cambridge University Press, 2001), Chap.~2.

\bibitem{carlip} S. Carlip, Rep. Prog. Phys. {\bf 64}, 885 (2001).

\bibitem{Isham} C. J. Isham, Structural Issues In Quantum Gravity. In: 
\textit{General Relativity and Gravitation: GR14}, E. Sorace, G. Longhi, L.
Lusanna and M. Francaviglia, eds. (World Scientific, Singapore, 1997).

\bibitem{HGRG} M. J. W. Hall, to appear in GRG; eprint gr-qc/0408098.

\bibitem{RBJP} M. Reginatto, Braz. J. Phys. {\bf 35}, 476 (2005).

\bibitem{MTW} C. W. Misner, K. S. Thorne and J. A. Wheeler, Gravitation
(Freeman, San Francisco, 1973).

\bibitem{zehkief} H.D. Zeh, {\it The physical basis of the direction of time} (Springer, Berlin, 1992), Chap.~6; C. Kiefer and H.D. Zeh, Phys. Rev. D {\bf 51} 4145 (1995).
\bibitem{gen}  The corresponding infinitesimal transformations are given by $P'=P+\epsilon\, \delta G/\delta S$, $S'=S-\epsilon\, \delta G/\delta P$, $\tilde{H}'=\tilde{H}+\epsilon\,\{\tilde{H},G\}$ 
\cite{goldstein}. Note that $P(x)=0$ implies a minimum at $x$ and hence that $\nabla P=0$ (for continuous $P$).  
 \bibitem{ang} Similarly, for three-dimensional particles the `ensemble angular momentum' $L:=\langle x\times\nabla S\rangle$ satisfies $\{L_i,L_j\}=\epsilon_{ijk}L_k$, and generates rotations of $P$ and $S$.  Either of $\tilde{H}_C$ or $\tilde{H}_Q$ (with $V\equiv 0$), together with $\Pi$, $L$ and $\Pi t-mX$, form a Poisson bracket representation of the Galilean group.



\end{thebibliography}
\end{document}